\documentclass[twocolumn]{aastex63}
\usepackage{graphicx}

\received{2 December 2020}
\revised{27 January 2021}
\accepted{29 January 2021}
\submitjournal{ApJ}
\acceptjournal{ApJ}

\shorttitle{Magnetic sheath around the 3C\,273 jet}
\shortauthors{Lisakov et al.}
\newcommand{\rmu}{\,rad\,m$^{-2}$\,} 
\newcommand{\rmumas}{\,rad\,m$^{-2}$\,mas$^{-1}$\,} 

\newcommand{\revi}[1]{#1} 

\usepackage{verbatim}

\begin{document}

\title{An oversized magnetic sheath wrapping around the parsec-scale jet in 3C\,273}

\correspondingauthor{Mikhail Lisakov}
\email{mlisakov@mpifr-bonn.mpg.de}

\author[0000-0001-6088-3819]{M. M. Lisakov}
\affiliation{Max-Planck-Institut f\"{u}r Radioastronomie, Auf dem H\"{u}gel 69, Bonn 53121, Germany}
\affiliation{Astro Space Center, Lebedev Physical Institute, Russian Academy of Sciences, Profsoyuznaya st., 84/32, Moscow, 117997, Russia}

\author[0000-0003-4540-4095]{E. V. Kravchenko}
\affiliation{Moscow Institute of Physics and Technology, Institutsky per. 9, Moscow region, Dolgoprudny, 141700, Russia}
\affiliation{Astro Space Center, Lebedev Physical Institute, Russian Academy of Sciences, Profsoyuznaya st., 84/32, Moscow, 117997, Russia}
\affiliation{INAF Istituto di Radioastronomia, Via P. Gobetti, 101, Bologna, 40129, Italy}

\author[0000-0002-9702-2307]{A. B. Pushkarev}
\affiliation{Crimean Astrophysical Observatory, Nauchny 298688, Crimea, Russia}
\affiliation{Astro Space Center, Lebedev Physical Institute, Russian Academy of Sciences, Profsoyuznaya st., 84/32, Moscow, 117997, Russia}
\affiliation{Moscow Institute of Physics and Technology, Institutsky per. 9, Moscow region, Dolgoprudny, 141700, Russia}

\author[0000-0001-9303-3263]{Y. Y. Kovalev}
\affiliation{Astro Space Center, Lebedev Physical Institute, Russian Academy of Sciences, Profsoyuznaya st., 84/32, Moscow, 117997, Russia}
\affiliation{Moscow Institute of Physics and Technology, Institutsky per. 9, Moscow region, Dolgoprudny, 141700, Russia}
\affiliation{Max-Planck-Institut f\"{u}r Radioastronomie, Auf dem H\"{u}gel 69, Bonn 53121, Germany}

\author[0000-0001-6214-1085]{T. K. Savolainen}
\affiliation{ Aalto University Department of Electronics and Nanoengineering, PL 15500, FI-00076 Aalto, Finland}
\affiliation{Aalto University Mets\"ahovi Radio Observatory, Mets\"ahovintie 114, FI-02540 Kylm\"al\"a, Finland}
\affiliation{Max-Planck-Institut f\"{u}r Radioastronomie, Auf dem H\"{u}gel 69, Bonn 53121, Germany}

\author[0000-0003-1315-3412]{M. L. Lister}
\affiliation{Department of Physics and Astronomy, Purdue University, 525 Northwestern Avenue, West Lafayette, IN 47907, USA}

\begin{abstract}

In recent studies, several AGN have exhibited gradients of the Faraday Rotation Measure (RM) transverse to their parsec-scale jet direction. Faraday rotation likely occurs as a result of a magnetized sheath wrapped around the jet. In the case of 3C\,273, using Very Long Baseline Array multi-epoch observations at 5, 8 and 15~GHz in 2009--2010, we observe that the jet RM has changed significantly towards negative values compared with that previously observed. These changes could be explained by a swing of the parsec-scale jet direction which causes \revi{synchrotron emission to pass through} different portions of the Faraday screen. 
We develop a model for the jet--sheath system in 3C\,273 where the sheath is wider than the single-epoch narrow relativistic jet. 
We present our oversized sheath model together with a derived wide jet full intrinsic opening angle $\alpha_\mathrm{int}=2.1^\circ$ and magnetic field strength $B_{||}=3$~$\mu$G and thermal particle density $N_\mathrm{e}=125~\mathrm{cm}^{-3}$ at the wide jet--sheath boundary 230~pc downstream (deprojected) from its beginning. Most of the Faraday rotation occurs within the innermost layers of the sheath. 
The model brings together the jet direction swing and long-term RM evolution 
and may be applicable to other AGN jets that exhibit changes of their apparent jet direction. 

\end{abstract}


\section{Introduction} 
\label{sec:intro}

General relativistic magnetohydrodynamic (GRMHD) simulations illustrate the key role of magnetic fields in the formation of relativistic jets in Active Galactic Nuclei \citep[AGN; e.g.][]{2001Sci...291...84M, 2004ApJ...605..656V, 2011MNRAS.418L..79T, 2014Natur.510..126Z}.
The topology and structure of these magnetic fields remains an active area of research.
The rotation of the central supermassive black hole twists magnetic field lines into a helical shape that can be revealed by observations of transverse Faraday rotation gradients.
GRMHD simulations also show that accreting  supermassive black hole systems naturally form a spine--sheath structure of the relativistic jets as a result of transverse velocity stratification in the outflows \citep[e.g.][]{2006MNRAS.368.1561M, 2018ApJ...868..146N}.
The sheath, which may be a magnetized screen surrounding the jet or the jet boundary layer, has been suggested as a plausible source of the Faraday rotation \citep[e.g.][]{ 2008ApJ...675...79A}.

Very Long Baseline Interferometry (VLBI) observations of parsec-scale AGN jets \citep[e.g.,][]{2003ApJ...589..126Z, 2004ApJ...612..749Z, 2009MNRAS.393..429O} reveal Faraday rotations greater than 45\degr{} and a linear dependence of the polarization angle with the wavelength squared.
This implies that most of Faraday rotation occurs within the thermal magnetized media located in close proximity to the jet \citep[e.g.,][]{2004ApJ...612..749Z}.
The detection of a Rotation Measure (RM) gradient across the jet  \citep{2002PASJ...54L..39A} in the quasar 3C\,273 (z=0.158; \citealt{1992ApJS...83...29S}), and subsequently in other AGN \citep[e.g.,][]{2014MNRAS.444..172G, 2017MNRAS.467...83K}, 
support the scenario that the sheath is threaded by a helical magnetic field.

\begin{figure}
\centering
\includegraphics[angle=-90,width=0.85\columnwidth]{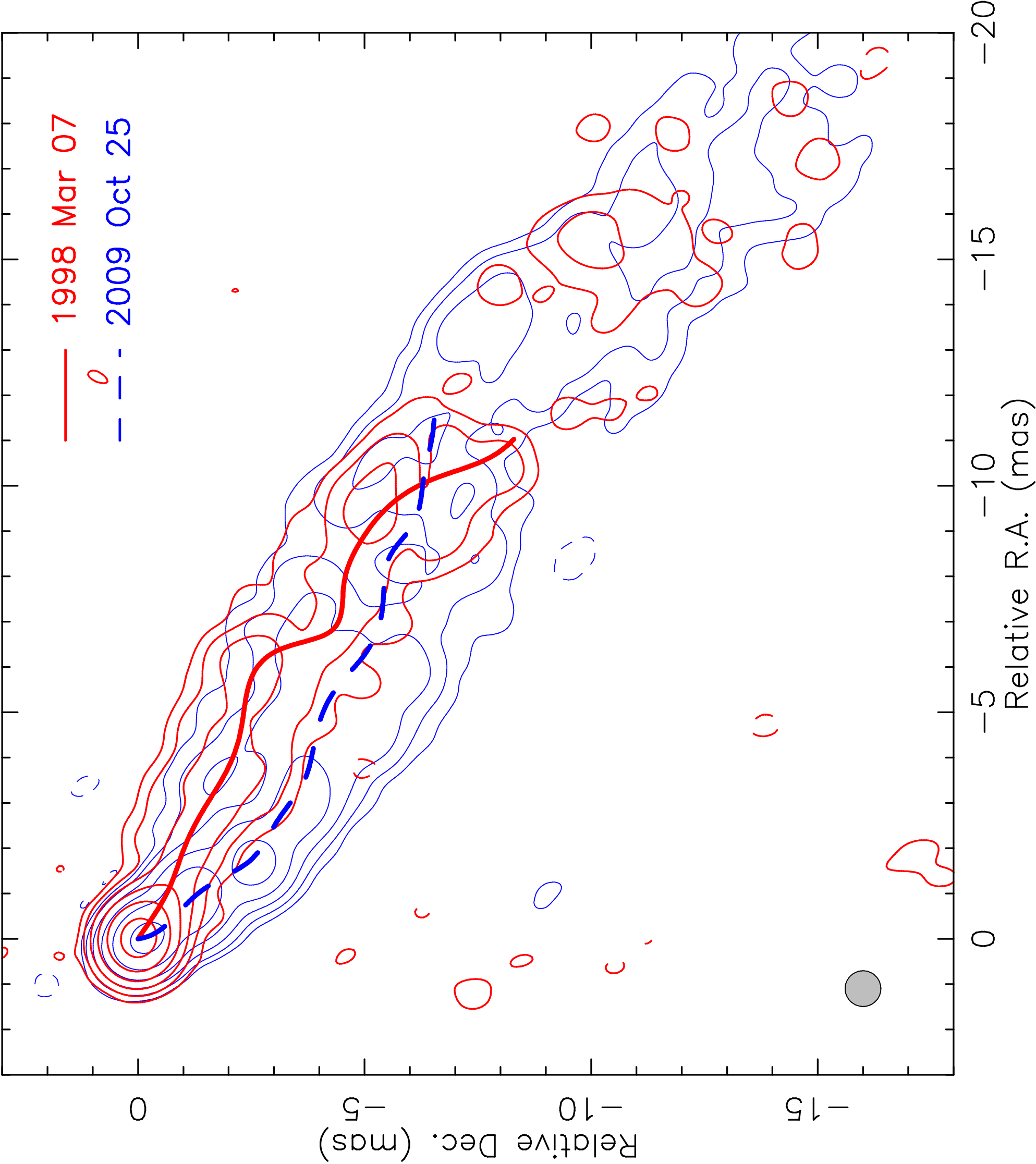}
    \caption{15\,GHz total intensity contours of 3C\,273 for epochs 1998 Mar 07 (red, taken from the MOJAVE database) and 2009 Oct 25 (blue, from present study). Both images were restored with the same, average equal-area circular Gaussian FWHM beam size of 0.8\,mas (shown in the bottom-left corner) for easier comparison. The images are aligned at the position of the Stokes $I$ peaks. \revi{Solid and dashed} lines mark the fitted jet ridge lines at the two epochs.}
\label{fig:jetpa}
\end{figure}

A number of studies have revealed that AGN jets change their direction on parsec-scales over time \citep[e.g.,][]{2006A&A...446...71S, 2013AJ....146..120L}, including 3C\,273 (Fig.~\ref{fig:jetpa}).
As different components are ejected in different directions into the jet, they fill in the full jet cross-section.
Therefore, the true jet width appears wider after stacking together observations performed over many years \citep{2017MNRAS.468.4992P,2020MNRAS.495.3576K}. Hereafter we call the latter a wide jet, while a single-epoch jet is called a narrow jet. 

3C\,273 (1226$+$023) has been extensively studied with polarimetric VLBI observations, and it remains the best-established case of a transverse RM gradient in AGN jets.
In this paper, we present a new multi-epoch high-resolution polarimetric study of 3C\,273, and Faraday rotation measure analysis, focusing on its long-term evolution, in order to find a connection between the apparent jet orientation and the values of the Rotation Measure. In Section~\ref{sec:data}, we describe our observations and data reduction steps. In Section~\ref{sec:slices}, we collect several transverse slice measurements from published research and compare them with our measurements. In Section~\ref{sec:discussion}, we present a model of an oversized sheath wrapping around the parsec-scale jet in 3C\,273 and discuss the geometry of the jet and physical parameters of the plasma in the sheath.

One milliarcsecond corresponds to a projected linear scale of 2.71\,pc at the redshift $z=0.158$ of 3C\,273, assuming $H_0 = 71$~km\,s$^{-1}$\,Mpc$^{-1}$, $\Omega_{\mathrm{m}} = 0.27$ , and  $\Omega_{\mathrm{\Lambda}} = 0.73$. For our calculations we use the viewing angle $\theta = 6\degr{}$ and the corresponding intrinsic full opening angle of the narrow single-epoch jet of 3C\,273 $\alpha_\mathrm{int}=1^\circ$ \citep{2017MNRAS.468.4478L}.

\section{Observations and data reduction} \label{sec:data}

The observations were performed with the 10-element Very Long Baseline Array (VLBA, project codes S2087A, B, C, E) on 2009 August 28, 2009 October 25, 2009 December 5, and 2010 January 26 at seven frequencies: 4.6, 5.0, 8.1, 8.4, 15.4, 23.8, 43.2~GHz with a bandwidth of 16~MHz for frequencies below 15.4~GHz and 32~MHz for 15.4~GHz and above.
For the aims of this study, we consider only the 4.6--15.4\,GHz frequency range, while detailed polarization analysis of the quasar in the full frequency range will be presented in a separate publication (Lisakov et al. in prep.).
Data reduction and calibration, as well as other techniques used for the analysis were discussed by \cite{2017MNRAS.468.4478L}. 

\begin{figure}
    \centering
    \includegraphics[width=0.29\textwidth,angle=270]{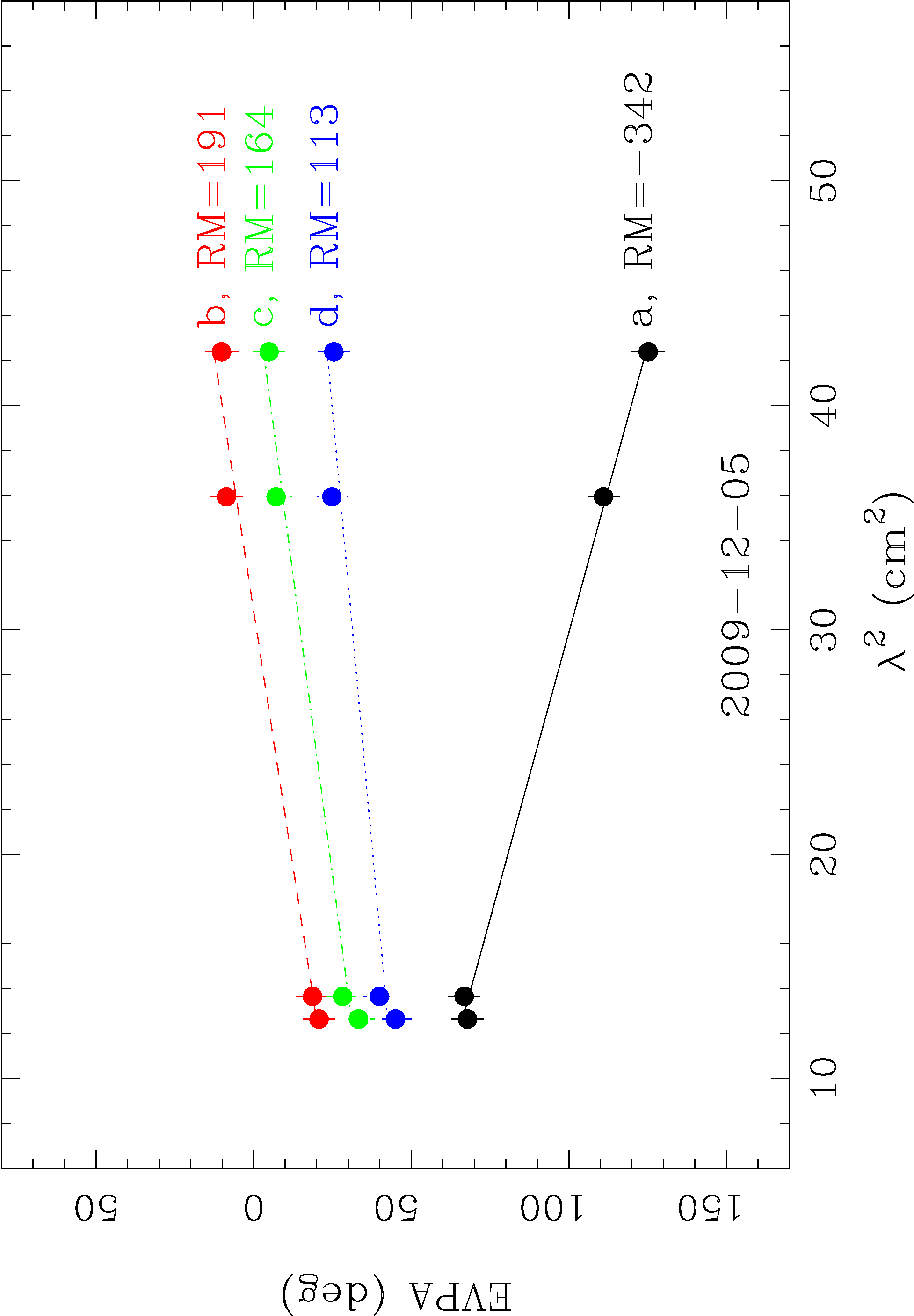}
    \caption{The plot of EVPA vs. $\lambda^2$ at locations in the jet shown in Figure~\ref{fig:low}, with the linear fits and the resultant RM values given in \rmu{}. Colours correspond to those used in Fig.~\ref{fig:low}.}
    \label{fig:evpafit}
\end{figure}

The polarization calibration was performed in the same manner as described in \citet{2017MNRAS.467...83K} and \citet{2017Galax...5...92K}, including instrumental polarization of the antennas that was determined with the task LPCAL in AIPS using 1308$+$326 as a calibrator.
An absolute orientation of the electric vector position angle (EVPA) was determined using the net polarization angle of 3C\,273, 3C\,279, OJ\,287, and 1308$+$326 provided by different monitoring programs\footnote{VLBA Polarization Calibration Resources \url{http://www.vla.nrao.edu/astro/calib/polar/}, UMRAO database \url{https://dept.astro.lsa.umich.edu/datasets/umrao.php}, and MOJAVE database \url{http://www.physics.purdue.edu/astro/MOJAVE/}.} with estimated uncertainty of 5\degr{} and 4\degr{} at 4.6--8.4 and 15.4\,GHz, respectively.

\begin{figure*}
    \centering
    \includegraphics[width=0.33\textwidth]{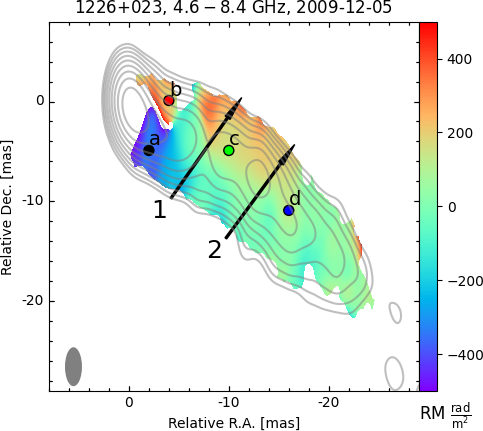}
    \includegraphics[width=0.305\textwidth]{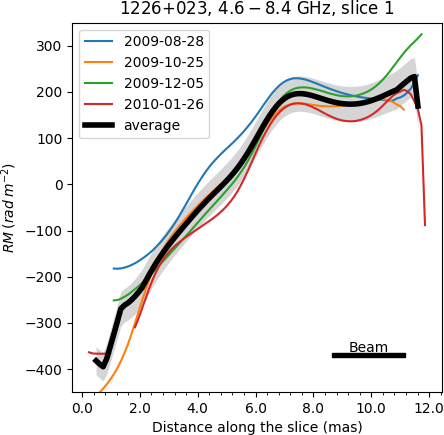}
    \includegraphics[width=0.295\textwidth]{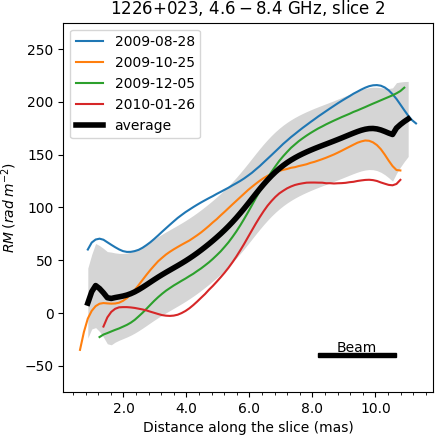}
    \caption{Left: 4.6--8.4~GHz Rotation Measure image (color) overlaid on the 4.6~GHz total intensity contours for 3C\,273 at epoch 2009~December~05. The FWHM beam size is indicated in the bottom left corner. Points indicate locations where the EVPA($\lambda^2$) data were extracted as shown in Fig.~\ref{fig:evpafit}. The location of the slices are displayed by arrows. The first slice (center) is taken at a distance 9~mas from the image center, as in \cite{2008ApJ...675...79A}.
    Right: a reference slice taken at 16~mas from the image center in the region that should not be affected by the jet direction change by 2009. In both slices, color lines represent measured RM values at each of the individual observations, while the thick black line is the average. The shaded area shows the average RM uncertainty. The beam projection FWHM on each slice is shown as a bar.}
    \label{fig:low}
\end{figure*}

Frequency-dependent map offsets were determined using 2D cross-correlation on optically thin portions of the jet  \citep{2017MNRAS.468.4478L} and were used to align images at the different frequencies. Average map offsets in the jet direction are 0.8~mas between 4.6 and 8.1~GHz and 0.6~mas between 8.1 and 15.4~GHz.
The Faraday rotation analysis was performed separately in two frequency ranges: low (4.6 -- 8.4~GHz) and middle (8.1 -- 15.4~GHz) in order to avoid smearing, which results from a convolution with larger beam sizes. 
Images at different frequencies were tapered to approximate the resolution of the lower frequency in the corresponding range.
All maps were convolved with the restoring beam averaged over four observations at 4.6 and 8.1\,GHz in corresponding frequency ranges.
We calculated the polarization errors according to \citet{2012AJ....144..105H}, and blanked all pixels whose polarized flux density did not exceed three times the polarization error.
The RM is defined as a linear slope of the EVPA$(\lambda ^2)$ dependence. Examples of EVPA$(\lambda ^2)$ fits for different parts of the jet are given in Fig.~\ref{fig:evpafit}. The solution of the n$\pi$-ambiguity problem and goodness of the $\lambda^2$-fit were determined by minimizing the reduced $\chi^2$. Accordingly, the pixels that had $\chi^2>5.99$ (which in our case of four data points and two degrees of freedom corresponds to the 95 per cent confidence level) were blanked.

\section{RM images and slices} \label{sec:slices}

\begin{figure*}
    \centering
    \includegraphics[width=0.33\textwidth]{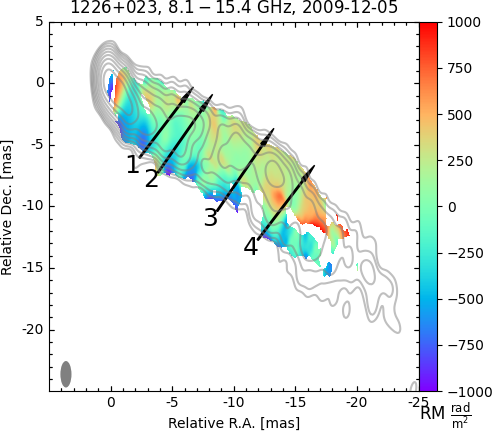}
    \includegraphics[width=0.30\textwidth]{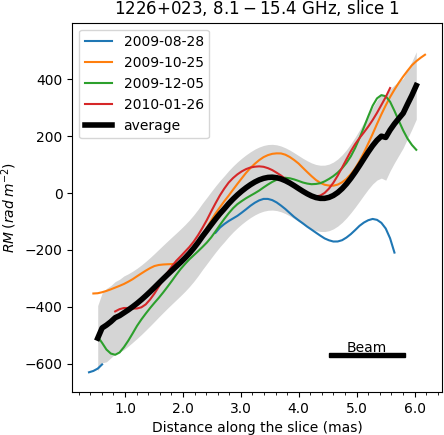}
    \includegraphics[width=0.30\textwidth]{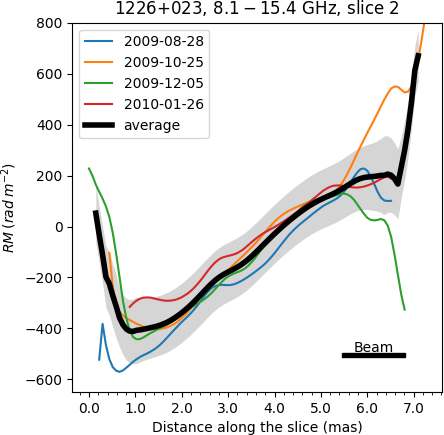}\\\vspace{10pt}
    \includegraphics[width=0.30\textwidth]{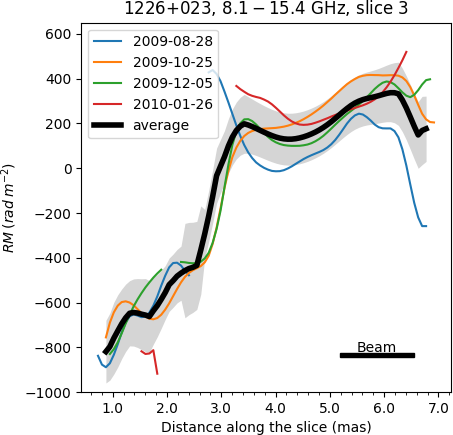}
    \includegraphics[width=0.30\textwidth]{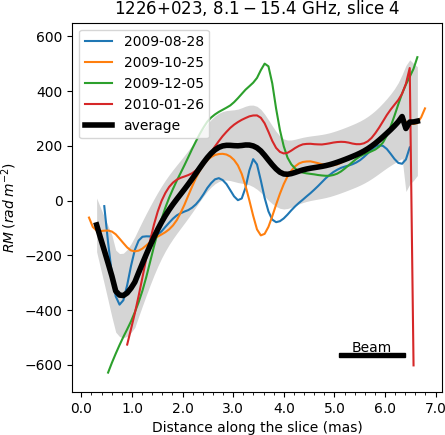}
    \caption{The same as in Fig.~\ref{fig:low}, but for the 8.1--15.4~GHz range, with the total contours from 8.1\,GHz at epoch 2009~Dec~05.}
    \label{fig:mid}
\end{figure*}

Faraday Rotation Measure images are presented in Fig.~\ref{fig:low} for the observations at 4.6--8.4\,GHz and in Fig.~\ref{fig:mid} for the 8.1--15.4\,GHz range.
Only a single-epoch image is presented, since the RM distribution throughout the jet is consistent over four epochs and does not change dramatically in regions downstream from the core. 
By the `core' or `apparent core', we mean the most north-eastern region of the jet in the VLBI images. We associate it with the apparent base of the jet and a surface with an optical depth $\tau\approx1$ \citep{1979ApJ...232...34B,1998A&A...330...79L}.
The overall RM distribution is smooth and shows persistent transverse gradients throughout the whole jet length, at least up to 57~pc in projection along the jet.

The absence of detected Rotation Measure close to the apparent core of 3C~273  is usually attributed to depolarization that happens in this region \citep[e.g.][]{2002PASJ...54L..39A, 2012JPhCS.355a2008H}.
We note that during our observations in 2009--2010 there was a major $\gamma$-ray flare detected in 3C~273 and a subsequent flare at radio-wavelengths \citep{2017MNRAS.468.4478L}. This is also supported by RadioAstron observations performed in 2012--2014 \citep{2016ApJ...820L...9K, 2017A&A...604A.111B} that show severe variability in the finest-scale structure of 3C~273 which is associated with the apparent core region. This variability together with the low level of polarized emission lead to blanking of the RM values close to the apparent jet base according to our criteria described in Section~\ref{sec:data}. 

To study the long-term evolution of the Rotation Measure and its gradient across the jet, we have collected previously published results on 3C\,273. 
\cite{2002PASJ...54L..39A} and \cite{2008ApJ...675...79A} reported a gradient of RM values taken across the jet at 4.6--8.6~GHz $9$~mas downstream from the apparent core. 
In their observations performed on 1995 December 9, 1995 November 22, and 2002 December 15, RM values across the jet from south to north have ranges $[130,370]$\rmu{} and $[170,480]$\rmu{}, which correspond to a gradient of $\approx 50$\rmumas{}. 
\cite{2005ApJ...626L..73Z} reported a much larger transverse gradient of 500~\rmumas{} taken slightly closer to the apparent core at $12-22$~GHz. 
Their observations were performed in 2000 January~27 and 2000 August~11, and also exhibit mostly positive RM values along the slice, $[150,2350]$\rmu{} and $[-50,1900]$\rmu{} at the two epochs, respectively. 

\cite{2018Galax...6....5W} compiled most of the RM measurements including those of \cite{2005PhDT........17C} performed in 1999--2000 at $8-22$~GHz. All RM values transverse to the 3C\,273 jet are positive, ranging within $[200,800]$~\rmu{} at the same location as in the previous works. RM gradients in these observations were $\approx 200$\rmumas{}.

A notable new result came from the research of \citet{2012AJ....144..105H} based on 8--15~GHz observations on 2006 March 9 and 2006 June 15, where the authors for the first time observed significant negative RM values, in contrast to earlier observations.
\citet{2012AJ....144..105H} report RM values across the jet ranging within $[-600,500]$\rmu{}, yielding a RM gradient of the same sign $\approx 180$\rmumas{}.

To compare with previous studies, we have analyzed transverse slices from our data taken at 9~mas \citep[as in][]{2008ApJ...675...79A} and 16~mas (control slice) from the apparent jet base at 4.6--8.4~GHz and 5.5~mas \citep[as in][]{2005ApJ...626L..73Z}, 7~mas \citep[as in][slice 2]{2012AJ....144..105H}, 13~mas \citep[as in][slice 3]{2012AJ....144..105H}, and 16~mas (control slice) at 8.1--15.4~GHz.

As is seen in Fig.~\ref{fig:low}~(slice~1) and Fig.~\ref{fig:mid}~(slice~1), closer to the apparent core, RM values across the jet are negative at the southern edge of the jet and positive at the northern edge. The range of values is $[-400,200]$\rmu{} at 4.6--8.4~GHz and $[-500,400]$\rmu{} at 8.1--15.4~GHz. The latter agrees well with the measurements of \citet{2012AJ....144..105H}. 
On the other hand, at 4.6--8.4~GHz, farther from the jet beginning, the RM values are positive throughout the whole cross-section of the jet, see Fig.~\ref{fig:low}~(slice~2). These positive RM profiles are more similar to the 4.6--8.6~GHz results of \citet{2008ApJ...675...79A}. 

At middle frequencies, transverse RM variations are present at every separation from the core, showing negative values at the southern edge and positive at the northern edge. We note, that due to our higher bandwidth and longer integration times we detect a broader RM distribution across the jet with respect to previous studies, hence more negative RM values are detected at the southern edge of the jet. 

\section{Discussion} \label{sec:discussion}

\subsection{RM variability}

With a large set of observations that span over 14 years and probe variability timescales down to several months, we can study both long-term and short-term evolution of the Rotation Measure values in the jet. We focus here on comparing transverse slices. 

Fig.~\ref{fig:low} and Fig.~\ref{fig:mid} clearly show that within the 5 months covered by our observations, the RM values across the jet at different distances from the apparent core do not change significantly. For the whole jet, net RM values change by no more than by 40\rmu{} at 4.6--8.4~GHz and by no more than 120\rmu{} at 8.1--15.4~GHz. These values do not exceed the uncertainty associated with the absolute EVPA calibration. 
Within the 5-month period covered by our observations in 2009--2010 the RM pattern in the jet of 3C~273 has remained largely constant, consistent with the measurements of \cite{2005ApJ...626L..73Z} over 6 months and three years. This is in contrast to the short timescale variability over three months as reported by \citet{2012AJ....144..105H}.

When considering variability on timescales of years, there are notable changes in the RM distribution in the jet of 3C\,273. \cite{2008ApJ...675...79A} reported on the small changes in the RM over 7 years. But a major change appears to have occurred some time between 2003 and 2006. Namely, at 7~mas downstream the jet, there are now negative values at the southern edge of the jet. This was first discovered by \citet{2012AJ....144..105H} and is fully confirmed with our observations at both 4.6--8.4~GHz and 8.1--15.4~GHz, see e.g. Fig.~\ref{fig:low}~(slice~1) and Fig.~\ref{fig:mid}~(slice~1).

Although there are alternative explanations of the long-term evolution of the RM values, we argue that anything outside the immediate jet vicinity fails to explain variability on the timescale of several years and large changes in RM, considering the high Galactic latitude $b=64.4\degr{}$ of 3C\,273 \citep{2009ApJ...702.1230T}.


\subsection{Geometry of the jet of 3C\,273} \label{geometry}

According to \cite{2013AJ....146..120L}, the narrow jet of 3C\,273 had abruptly changed its apparent direction by $20^{\circ}$ in 2003. 
Fig.~\ref{fig:jetpa} shows these structural variations in the quasar jet in application to the observations of 1998 March 7 and 2009 October 25 at 15~GHz. 
Assuming a constant viewing angle of the overall quasar jet axis of $\theta = 6^{\circ}$ \citep{2017MNRAS.468.4478L}, this translates into an intrinsic jet direction change of $2^{\circ}$, larger than the intrinsic opening angle measured at a single epoch $\alpha_{\mathrm{int}} = 1.1^{\circ}$ \citep{2017MNRAS.468.4478L}. 

With a median apparent jet speed of $0.8$~mas~yr$^{-1}$ \citep{2019ApJ...874...43L}, structural changes should affect the jet by 2010 within about 7~mas from its apparent beginning, consistent with Fig.~\ref{fig:jetpa}. We also note that some moving features in the jet of 3C\,273 have larger apparent speeds, up to $1.5$~mas~yr$^{-1}$, which could potentially double the affected jet region \citep{2006A&A...446...71S}.

\begin{figure}
    \centering
    \includegraphics[width=0.45\textwidth]{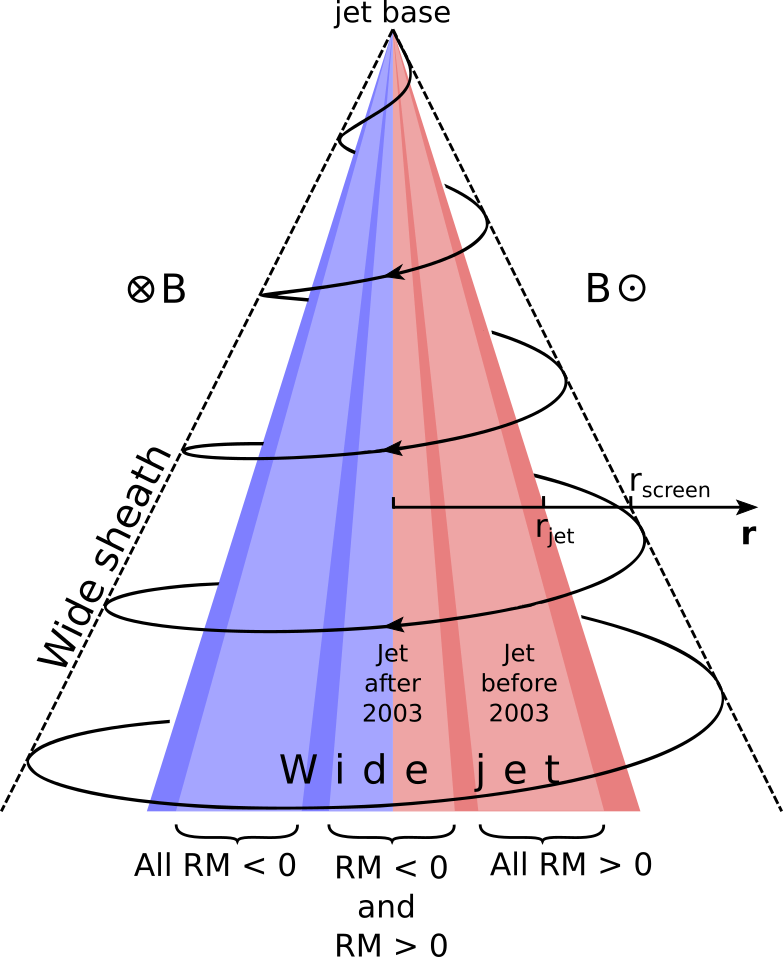}
    \caption{Sketch of a sheath wrapped around the wide jet. Subject to the single-epoch relativistic jet orientation, \revi{synchrotron emission of the jet passes through} different parts of the sheath. RM values are all positive before 2003 (rightmost jet position) and cross zero after 2003 (middle jet position). If the jet of 3C\,273 turns more to the south (to the leftmost position in this sketch), only negative RM values will be observed.}
    \label{fig:model}
\end{figure}

\subsection{Wide Faraday sheath in 3C\,273} \label{sec:model}

We suggest that the Faraday rotating medium consists of an oversized sheath wrapped around the wide jet instead of being tightly connected with the narrow, single-epoch jet as shown in Fig.~\ref{fig:model}. The inner opening angle of the sheath is therefore larger than the opening angle of the narrow single-epoch jet that wiggles within the inner opening angle of the oversized sheath. 
In this scenario, there is a misalignment between the visible narrow relativistic jet axis and the symmetry axis of a broader conical sheath which is threaded with toroidal magnetic field.
Therefore, due to the swing in the parsec-scale jet direction occurring in 2003, \revi{synchrotron radiation passed through} different regions of the Faraday screen and different RM values were observed before and after this change. This possibility was previously considered to explain some RM changes in 3C~273 \citep{2012JPhCS.355a2008H} and 3C~120 \citep{2011ApJ...733...11G}.

There are several possible ways to verify this scenario. Firstly, the screen is detached from the relativistic jet and most probably has a much slower velocity. Therefore, it should evolve relatively slowly. Indeed, for our data, we have measured the jet net RM changes between observations in 2009--2010 as: $-32$, $-3$, $-25$~\rmu\ at 4.6--8.4~GHz and $114$, $-34$, $63$~\rmu{} at 8.1--15.4~GHz. These values are within the uncertainties associated with the EVPA calibration.
Secondly, should the rotating medium constitute a sheath around the narrow relativistic jet with thickness comparable to the jet cross-section, it will inevitably be destroyed by the relativistic jet once it changes direction. However, we do not detect any traces of such an interaction. 

We note that \citet{2019ApJ...871..257P} present RM images of the jet in the radio galaxy M87 within the Bondi radius and conclude that the dominant source of the observed significantly high RMs are hot winds. The latter are non-relativistic, moderately magnetized gas outflows that surround the highly magnetized jet of M87 and are threaded by toroidally-dominated magnetic fields.
But M87 shows no RM gradients across its jet.
\citet{2019ApJ...871..257P} suggest that this is due to a misalignment between the jet axis and the symmetry axis of the toroidal field loops in the Faraday screen, therefore the RM gradients are not observed in M87 since only a portion of the sheath is illuminated.
Although the matter which forms the sheath in M87 and in 3C\,273 may have a different origin, observational properties for both sources are well explained within the presented model. 

\revi{Our model of an oversized sheath explains the long-term RM variability in the inner parts of the jet with different apparent directions of the relativistic jet before and after 2003 and hence the jet synchrotron emission passing through different parts of the Faraday screen.} We can also predict that if the jet direction changes farther to the south, mostly negative RM values will be observed. 

\subsection{Screen parameters} \label{sec:screen}

To estimate parameters of the Faraday screen we have developed a simple model. The wide jet with a radius $r_\mathrm{jet}$ is wrapped by a hollow conical sheath with inner radius  $r_\mathrm{jet}$ and outer radius  $r_\mathrm{screen}$, see Fig.~\ref{fig:model}. The toroidal component of the magnetic field in the sheath decays with perpendicular distance from the wide jet axis as $B_{||}=B_1 r^{-1}$ and the thermal particle density declines as $N_\mathrm{e} = N_1 r^{-2}$ where $B_1$ and $N_1$ are taken at 1~pc from the wide jet axis. 
After integrating \revi{$RM \propto \int B_{||} N_\mathrm{e} dl $} over different lines of sight \revi{perpendicular to the apparent jet direction}, we obtained the distribution of the RM values across the jet width. We then convolved these with a \revi{Gaussian} restoring beam assuming five beams across the jet \revi{to resemble the data at 4.6--8.4~GHz}. \revi{This simple model is designed to account only for smooth and monotonic distribution of RM across the jet and give a hint of typical parameters of the medium in the real sheath.}

\begin{figure}
    \centering
    \includegraphics[width=0.45\textwidth]{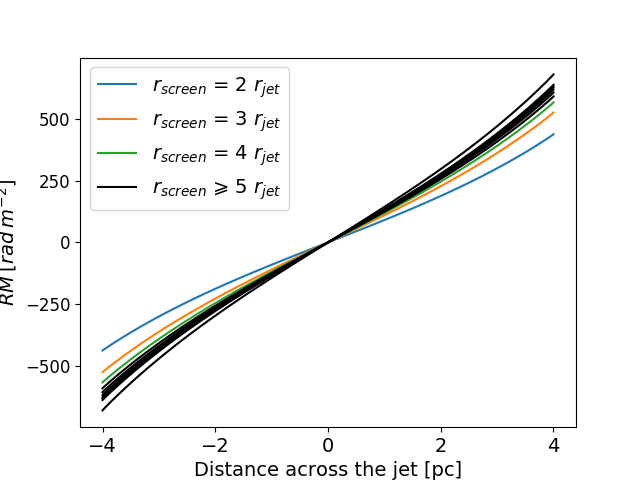}
    \caption{Model RM values across the jet for the 4.6--8.4~GHz range at 9~mas from the core. $B_1 = 12$~$\mu G$ and $N_1 = 2000~\mathrm{cm}^{-3}$ were chosen to obtain the total range $[-500, 500]$\rmu{} of observed RM values for $r_\mathrm{screen} = 3\:r_\mathrm{jet}$. The family of black curves represents $r_\mathrm{screen}$ values ranging from $ 5\:r_\mathrm{jet}$ to $100\:r_\mathrm{jet}$. 
    }
    \label{fig:modelrm}
\end{figure}

For any $r_\mathrm{screen}$, the inner layers of the screen, up to $r_\mathrm{screen} = 3\times r_\mathrm{jet}$, contribute 75\% of the observed RM \revi{as expected from the model}, see Fig.~\ref{fig:modelrm}. This supports our initial assumption that the sheath would be disrupted by the jet changing its direction if this sheath is located in the immediate vicinity of the narrow relativistic jet. \revi{Moreover, the predicted RM distribution across the jet is well described by a straight line.}

\begin{figure}
    \centering
    \includegraphics[width=0.45\textwidth]{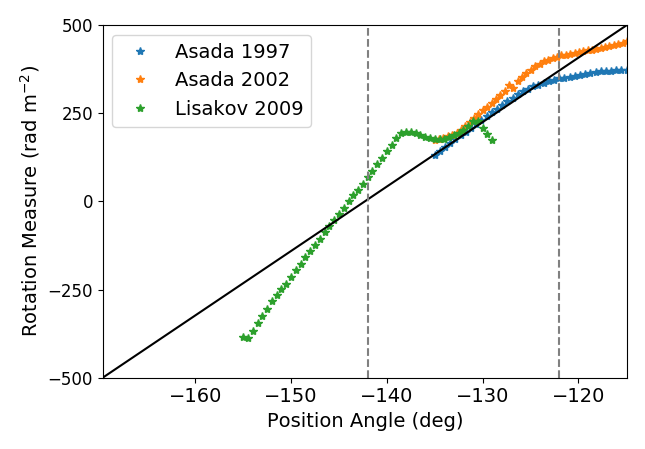}
    \caption{The RM values at 4.6--8.4~GHz taken at 9~mas from the apparent jet beginning transverse to its direction. Blue and orange points represent two epochs, 1997 and 2002,  reported by \cite{2008ApJ...675...79A}, while green points are taken from our observation on Dec~05, 2009. The two vertical dashed lines show the average narrow jet position angle before (at $-122^\circ$) the major jet swing in 2003, and after (at $-142^\circ$). The black diagonal line represent a linear fit to the combined data set.}
    \label{fig:glued_rm}
\end{figure}


One may assume that the maximum value of the RM across the jet at 9~mas from the apparent core measured by \cite{2002PASJ...54L..39A} was 500\rmu{} and that the minimum value, not observed yet, would be $-500$\rmu. Then with our measurements of RM values across the jet at 4.6--8.4~GHz taken at the same distance, we estimate a total wide jet intrinsic opening angle of $\alpha_\mathrm{int} = 2.1\degr{}$, twice as large as single-epoch estimates \citep{2017MNRAS.468.4478L}. To perform this, we combine transverse RM distributions, taken at different epochs, with the narrow jet position angle changes taken into account, as is shown in Fig.~\ref{fig:glued_rm}. The apparent angular width of the RM distribution at a single epoch is estimated as the width of the transverse jet intensity profile taken at the 1\% level. Afterwards, the combined transverse RM distribution was fitted with a linear function\revi{, in accordance with the model}. The total range of jet position angles, required to obtain a $[-500, 500]$\rmu{} range of the RM, is 55$^{\circ}$ which corresponds to the intrinsic opening angle of the wide jet $\alpha_\mathrm{int} = 2.1\degr{}$.

With this estimate, at 9~mas from the apparent core, the total width \revi{of the wide jet} is 8~pc\revi{, $r_\mathrm{jet}=4$~pc}. \revi{As we found for an arbitrary large $r_\mathrm{screen}$, the inner layers of the sheath up to $r = 3r_\mathrm{jet}$ are responsible for 75\% of the rotation. Hence we assumed $r_\mathrm{screen} = 3 r_\mathrm{jet}$.} The range of RM values $[-500,500]$ then requires $B_1\times N_1 = 24000\:\mathrm{\mu G}\:\mathrm{cm}^{-3}$ for a path length $2r_\mathrm{jet} = 8$~pc through the sheath. Clearly, $B_1$ and $N_1$ can not be calculated independently within this analysis. At the same time, the line-of-sight path length through the screen could not change the estimates of $B_1\times N_1$ by more than a factor of two within our model, for any reasonable screen width. With $r_\mathrm{jet}=4$~pc,  $B_{||}\times N_\mathrm{e} = 375\:\mathrm{\mu G}\:\mathrm{cm}^{-3}$ at the wide jet -- wide sheath boundary. For assumed \revi{magnetic field component parallel to the line of sight} $B_{||}=3\:\mathrm{\mu G}$ \revi{at the inner edge of the sheath}, the thermal particle density is $N_\mathrm{e}=125\:\mathrm{cm}^{-3}$.


With the same logic applied to the 8.1--15.4~GHz, a range of RM values $[-2000,2000]$ could be achieved at 5.5~mas from the jet beginning with $B_1\times N_1 = 37500\:\mu G\:\mathrm{cm}^{-3}$, as obtained from the model with path length of $2r_\mathrm{jet} = 5$~pc. $B_1 \times N_1$ is 25\% higher than that at 9~mas, what is expected given the lower distance to the jet beginning. 
\revi{As is apparent from Fig.~\ref{fig:mid}, transverse RM distributions observed at higher frequencies in general show more complex shape and thus require a more elaborate model to describe them. This is a scope of future work that will allow a better estimate of the Faraday screen parameters. }

It is important to note that using the core shift analysis \citep{1998A&A...330...79L} and estimates of the magnetic field $B_0=0.47$~G and relativistic electron density $N_0=1500$~cm$^{-3}$ at the jet beginning \citep{2017MNRAS.468.4478L}, the estimated values in the jet are $B_{140} = 3\:\mathrm{\mu G}$, $N_{140} = 8\times 10^{-2}\:\mathrm{cm}^{-3}$ at 5.5~mas (corresponding to 140~pc, deprojected) and $B_{230} = 2\:\mathrm{\mu G}$, $N_{230} = 3\times 10^{-2}\:\mathrm{cm}^{-3}$ at 9~mas(230~pc, deprojected) from the apparent jet base.

\begin{figure}
     \centering
     \includegraphics[width=0.45\textwidth,angle=0]{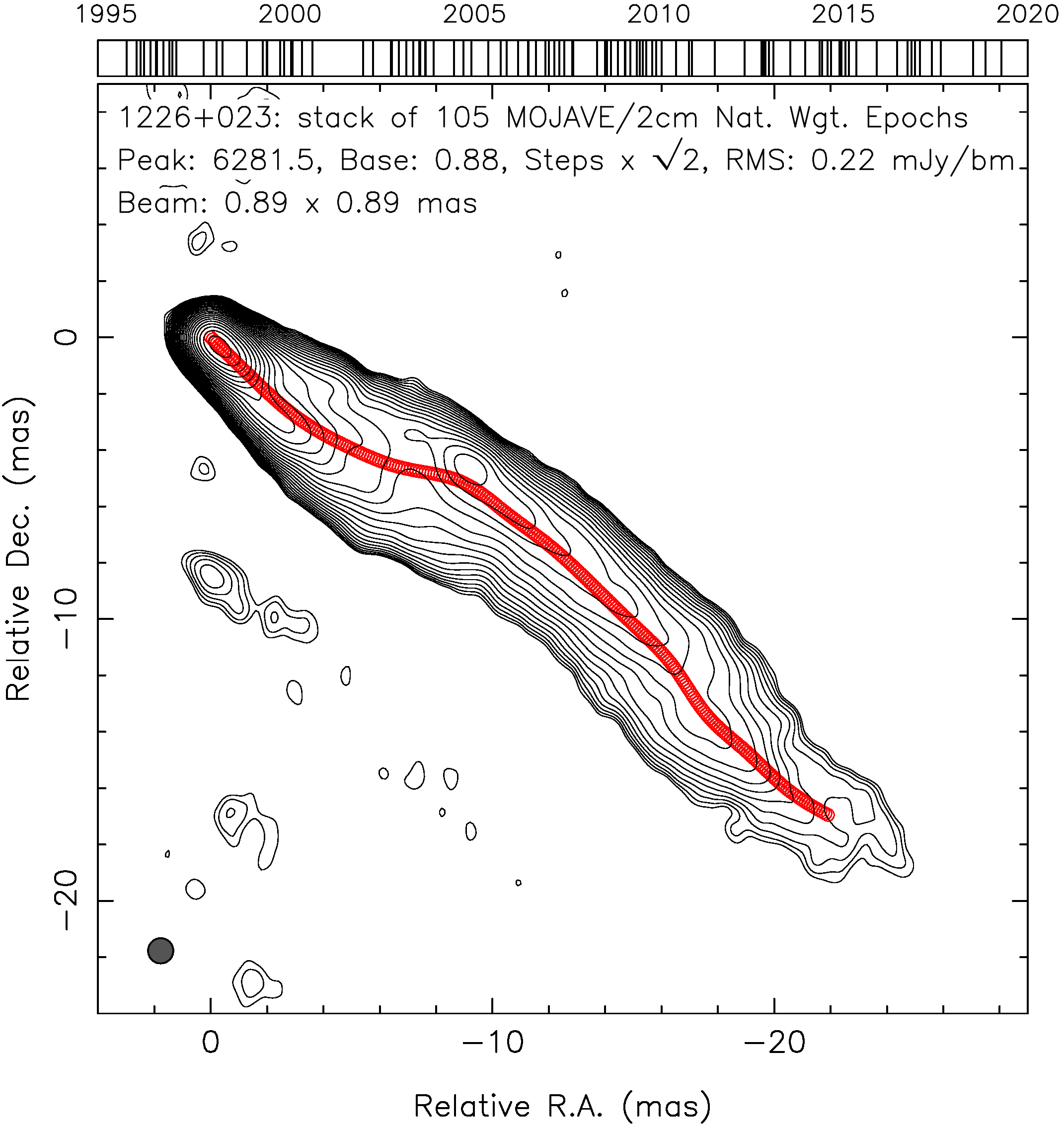}
     \caption{15~GHz naturally weighted stacked \textsc{clean} image of 3C\,273 based on 105 VLBA epochs from the MOJAVE program \citep{2019ApJ...874...43L}. The contours are shown at increasing powers of $\sqrt{2}$ times the base contour level of 0.88 mJy\,beam$^{-1}$. The restoring beam dimensions are plotted as a shaded circle in the lower left corner. The constructed total intensity ridge line is depicted in red. A wedge indicating observing epochs (vertical ticks) used for producing the stacked image is shown on top.}
     \label{fig:stack}
\end{figure}

\subsection{The jet opening angle derived from stacked images} \label{sec:stack}

\begin{figure}
    \centering
    \includegraphics[width=0.45\textwidth,angle=0]{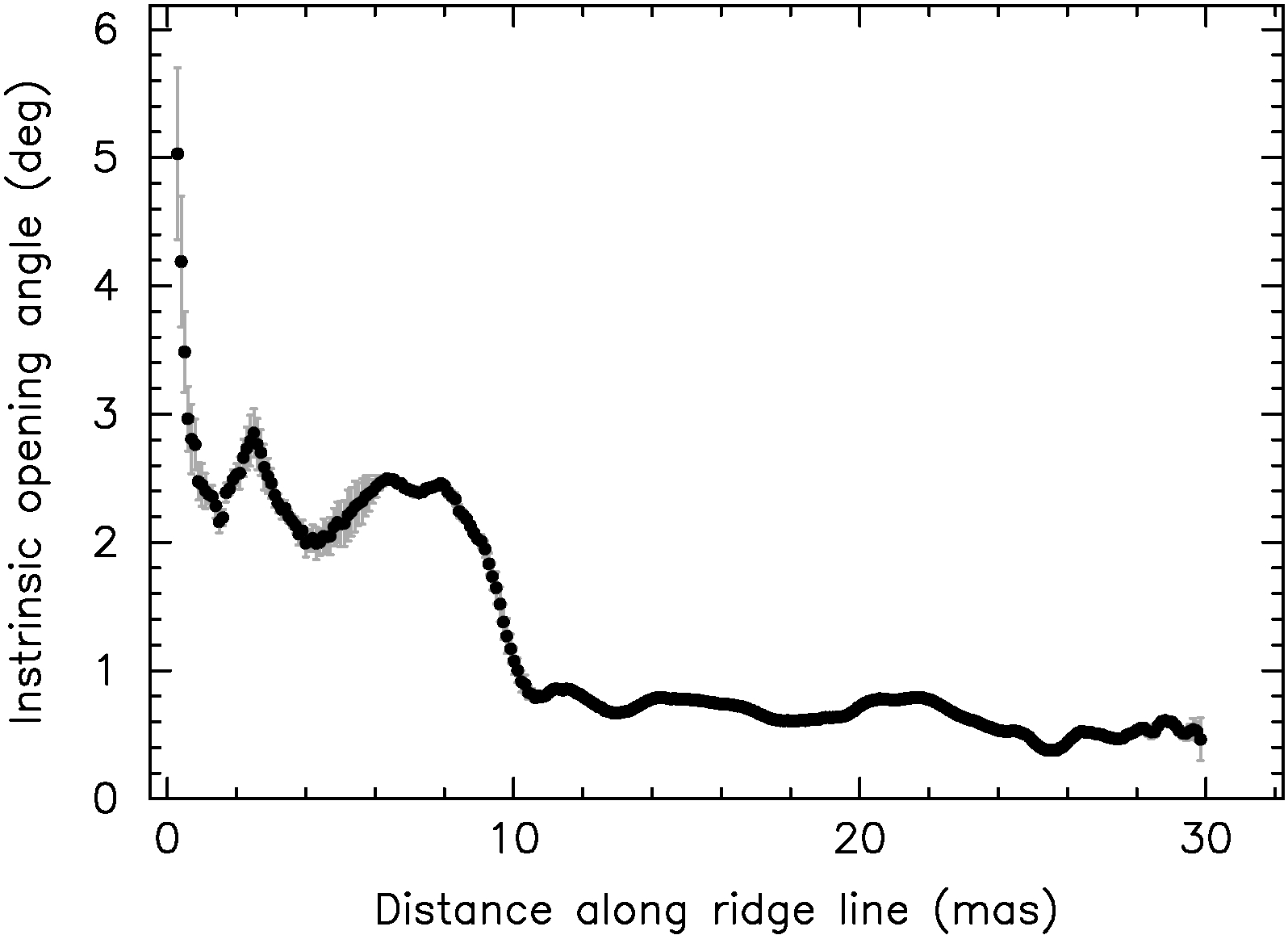}
    \caption{Intrinsic opening angle of the jet derived from stacked image in Fig.~\ref{fig:stack} as a function of the core separation measured along the ridge line.}
    \label{fig:ioa}
\end{figure}

We produced a stacked map of 3C\,273 in total intensity at 15~GHz (Fig.~\ref{fig:stack}) using public data from the MOJAVE database \citep{2019ApJ...874...43L}. According to discussion in Section~\ref{sec:model}, this map represents the wide jet in 3C\,273. In total, 105 epochs ranging from 1995 July 28 to 2019 April 15 were stacked together. Each single-epoch image was convolved with the same median circular beam and shifted such that the 15~GHz VLBI core is placed in the phase center. The core position \citep{2019ApJ...874...43L} was determined from source structure modeling performed in the Fourier plane with the procedure {\it modelfit} from the Caltech \textsc{Difmap} package \citep{difmap}. We fitted a ridge line to the image and measured the apparent wide jet opening angle $\alpha_\mathrm{app}$ following the procedure described in \cite{2017MNRAS.468.4992P}. The intrinsic wide jet opening angle, calculated as $\tan(\alpha_\mathrm{int}/2)=\tan(\alpha_\mathrm{app}/2)\sin\theta$, is shown in Fig.~\ref{fig:ioa}. The intrinsic opening angle of the wide jet is $\alpha_\mathrm{int}\approx2^\circ$ up to the distances of about 9~mas from the core along the ridge line, in noticeable agreement with the estimate made above under the assumption on the minimum Rotation Measure $-500$\,rad\,m$^{-2}$.

We also note that high resolution space VLBI images of 3C~273 obtained with RadioAstron at both 1.6~GHz and 4.8~GHz  and presented recently by \cite{bruni_radioastron_2021} show a clear evidence for the new established direction of the jet in 3C~273.

\section{Summary} \label{sec:summary}

We present new high-resolution images of the Faraday Rotation Measure in the parsec-scale jet of 3C\,273, see Section~\ref{sec:slices}, based on multifrequency VLBA polarimetric observations at 4.6--15.4~GHz. From the comparison of our results with the previous studies, we confirm the existence of transverse RM gradients in the jet of  3C\,273 out to at least 57~pc projected distance ($500$~pc deprojected), and detect a change of RM towards negative values compared with observations performed before 2003.

In order to coherently explain the change of the jet position angle together with the change of the Faraday Rotation Measure across the jet, in Section~\ref{sec:model} we develop a model of an oversized sheath that is wrapped around the wide jet in 3C\,273. The sheath does not show any rapid variability and most likely is disconnected from the relativistic jet. Single-epoch images of 3C\,273 reveal only a portion of the wider jet. After the parsec-scale jet direction swing in 2003, different lines-of-sight through the sheath are sampled, resulting in different observed Faraday Rotation Measure values. We predict that mostly negative RM values will be observed if the jet turns even more southward. 
Within our model we estimate the total wide jet intrinsic opening angle $\alpha_\mathrm{int} = 2.1\degr$, based solely on the RM measurements across the jet. Most of the rotation occurs within a layer of $3\times r_\mathrm{jet}$ width, see Section~\ref{sec:screen}. In this scenario, we estimate $B=3$~$\mu$G and $N=125$~$\mathrm{cm}^-3$ at the jet--sheath border at 9~mas (230~pc, deprojected) from the apparent core at 5~GHz.

We produced a stacked image of 3C\,273 in total intensity at 15~GHz using VLBA data at 105 epochs of observations performed during 24 years mainly within the MOJAVE program and its predecessor, the 2\,cm VLBA survey, see Section~\ref{sec:stack}. The intrinsic wide jet opening angle derived from the transverse cuts is about $5^\circ$ at the jet region near the 15~GHz core. The outflow quickly collimates to $\alpha_{\text{int}}\approx2^\circ$. This regime holds within distances from about 1 to 9~mas from the core, further supporting our model.

\acknowledgments
\revi{The authors thank the anonymous referee} and  Nicholas MacDonald for thoroughly reading the manuscript and providing useful comments. We thank Elena Nokhrina for a fruitful discussion while preparing this manuscript. This study has been supported in part by the Russian Science Foundation project 20-72-10078.
TS was partly supported by Academy of Finland projects 274477 and 315721.
This research has made use of data from the MOJAVE database that is maintained by the MOJAVE team \citep{2018ApJS..234...12L}. The MOJAVE program is supported under NASA-Fermi grant 80NSSC19K1579. 
This research has made of use of VLBA, which is run by the National Radio Astronomy Observatory is a facility of the National Science Foundation operated under cooperative agreement by Associated Universities, Inc. This work made use of the Swinburne University of Technology software correlator \citep{2011PASP..123..275D}, developed as part of the Australian Major National Research Facilities Programme and operated under licence.
\facilities{VLBA}
\software{AIPS \citep{AIPS}, Difmap \citep{difmap}, astropy \citep{2013A&A...558A..33A}, WebPlotDigitizer \url{https://automeris.io/WebPlotDigitizer}}

\bibliography{main}
\bibliographystyle{aasjournal}
\end{document}